%% file: ms.tex
\documentclass{article}

\usepackage{arxiv}
\usepackage[style=authoryear,citestyle=authoryear]{biblatex}
\usepackage[utf8]{inputenc} 
\usepackage[T1]{fontenc}    
\usepackage{hyperref}       
\usepackage{url}            
\usepackage{booktabs}       
\usepackage{amsfonts}       
\usepackage{nicefrac}       
\usepackage{microtype}      
\usepackage{lipsum}
\usepackage{graphicx}

\usepackage{paralist}

\usepackage{soul,xcolor}
\usepackage[htt]{hyphenat}
\definecolor{light-gray}{gray}{0.9}
\sethlcolor{light-gray}
\newcommand{\qs}[3]{`\emph{#1 [#2]}#3'}

\newcommand{\qt}[1]{`{\it #1}'}

\title{On the Social and Technical Challenges of Web Search Autosuggestion Moderation}

\author{
  Timothy J. Hazen\thanks{Correspondence may be directed to the first two authors at \href{mailto:tj.hazen@microsoft.com}{tj.hazen@microsoft.com} and \href{mailto:aloltea@microsoft.com}{aloltea@microsoft.com}.}\\
  Microsoft \\
  Cambridge, MA, USA \\
   \And
  Alexandra Olteanu$^*$ \\
  Microsoft \\
  New York City, NY, USA \\
  \AND
  Gabriella Kazai \\
  Microsoft \\
  Cambridge, UK \\
  \And
  Fernando Diaz\\
  Microsoft \\
  Montreal, Canada \\
  \And
  Michael Golebiewski\\
  Microsoft \\
  Bellevue, WA, USA \\
}

\begin{document}

\maketitle

\setcounter{footnote}{0}

\begin{abstract}
  Past research shows that users benefit from systems that support them in their writing and exploration tasks. 
  The autosuggestion feature of Web search engines is an example of such a system: It helps users in formulating their queries 
  by offering a list of suggestions as they type.
  Autosuggestions are typically generated by machine learning (ML) systems trained on a corpus of search logs and document representations. 
  Such automated methods can become prone to issues that result in problematic suggestions that are biased, racist, sexist or in other ways inappropriate. 
  While current search engines have become increasingly proficient at suppressing such problematic suggestions, there are still persistent issues that remain.
  In this paper, 
  we reflect on past efforts and on why certain issues still linger by covering explored solutions along a prototypical pipeline for 
  identifying, detecting, and addressing problematic autosuggestions.  
  To showcase their complexity, we discuss 
  several dimensions of problematic suggestions, 
  difficult issues along the pipeline, and 
  why our discussion applies to the increasing number of applications beyond web search that implement similar textual suggestion features.    
  By outlining persistent social and technical challenges in moderating web search suggestions, we provide a renewed call for action. 
\end{abstract}



\input{intro-background.tex}

\input{characterization.tex}

\input{discovery_detection.tex}

\input{challenges.tex}

\input{discussion.tex}

\section*{Acknowledgments}

The authors have benefited from collaborations and interactions with a range of colleagues that deserve recognition. Thus, the authors wish to provides thanks for discussions and feedback to Eugene Remizov, Joshua Mule, Jose Santos, Balakrishnan Santhanam, Abhigyan Agrawal, Swati Valecha, Harish Yenala, Shehzaad Dhuliawala, Zhi Liu, Keith Battocchi, Toby Walker, Molly Shove, Marcus Collins, Mohamed Musbah, and Luke Stark.

\printbibliography

\end{document}

%% file: intro-background.tex
\section{Introduction}

\begin{figure}
  \begin{center}
  \includegraphics[width=0.35\linewidth, trim={0cm 1.2cm 0cm 0cm},clip]{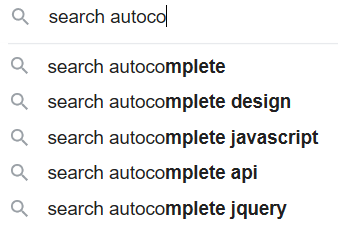}
  \end{center}
  \caption{Top 4 example suggestions provided by Google's autocomplete for the typed query prefix "search autoco".}
  \label{fig:AS-Example}
\end{figure}

Web search autosuggestion (often also referred to as autocompletion) is a feature enabled within the search bar of Web search engines such as Google~\parencite{Google-Autocomplete-Blog-2018} and Bing~\parencite{Bing-Autosuggest-Blog-2013a,Bing-Autosuggest-Blog-2013b}. 
It provides instant suggestions of complete queries based on the partial query entered by a user.\footnote{We use the term autosuggestion instead of autocompletion as suggestions are not restricted to completions of a partial query, but can also allow alternative wording.} 
Such query suggestions can not only help users to complete their queries with fewer keystrokes, but it can also aid them in avoiding spelling errors and, in general, guide users in better expressing their search intents.
Figure~\ref{fig:AS-Example} shows an example set of suggestions by a search engine for the query prefix "search autoco." 
Autosuggestion systems are typically designed to predict the most likely intended queries given the user's partially typed query, where the predictions are primarily based on 
frequent queries mined from the search engine's query logs~\parencite{Cai-FTIR-2016}.

Despite its advantages, there are also pitfalls that can accompany this search feature. 
Since the suggestions are derived from search logs, they can, as a result, be directly influenced by the search activities of the search engine's users.  
While this activity usually involves benign informational requests, it can also include queries that others may view as loathsome, prurient, or immoral.  
If left unchecked, a wide variety of problematic queries may be mined from the logs including, but not limited to, queries that provide exposure to violent, gory, or sexually explicit content; promote behavior that is harmful, hateful, or illegal; or wrongfully defame people or organizations. 
The feature could also suggest more pernicious queries that expose undesirable biases and stereotypes, or spread misinformation; among many other issues.

The dependence on query logs also leaves the autosuggestion feature susceptible to manipulation through adversarial attacks.  
By coordinating the submission of a large number of specific queries, it is possible 
for an outside entity
to directly manipulate the list of suggestions shown for targeted query prefixes \parencite{Starling-13}. 
Two common scenarios where this happens are online promotion campaigns and intentional "trolling." 
This issue is often enabled and further exacerbated by the so called {\em data voids}---topics searched in the past by little to no users and that reputable sources do not cover~\parencite{golebiewskid}---due to a lack of competing queries for rare query prefixes (\S\ref{sub:detection_issues}).

The presence of highly offensive or derogatory suggestions within prominent search engines has not gone unnoticed. 
Numerous negative stories in the news media have highlighted problematic suggestions made by the various search engines~\parencite{Elers-Te-Kahaoa-2014,Lapowsky-Wired-2018, Hoffman-How-To-Geek-2018,Chandler-The-Sun-2018}. 
Real search autosuggestions were, for instance, also used in a high profile 2013 ad campaign by UN Women to highlight the ``widespread prevalence of sexism and discrimination against women'' \parencite{UN-Women-AD-Campaign-2013}. 
The campaign highlighted examples like 
\qs{women shouldn't}{have rights}{,} 
\qs{women cannot}{be trusted}{,} 
\qs{women should}{be slaves}{} (where the part in square brackets represents a search suggestion that completes the partially typed query preceding it).\footnote{We are using this notation throughout the paper, i.e., `query prefix [suggestion]'} 
Similar issues have also been highlighted by legal complaints concerned with suggestions allegedly defaming an individual or an organization (e.g., a suggestion implying the plaintiff is a `scam' or a `fraud') or promoting harmful illicit activities (e.g., a suggestion pointing to pirated versions of a plaintiff's content)~\parencite{Ghatnekar-LLAELR-2013,Cheung-SSRN-2015,Karapapa_IJILT_2015}.

Sustained efforts to address these issues are critical to minimizing harms and maintaining public trust.
Such efforts to reduce problematic web search suggestions have resulted in noticeable improvements in recent years, though problems do remain. 
The actual mechanisms being used by particular search engines, however, have mostly remained hidden from the public view to avoid giving adversaries sufficient information to circumvent them.  
Thus, studies of issues surrounding autosuggestions tend to come from outside organizations that externally probe search engines using tools such as Google Trends\footnote{\url{https://trends.google.com/trends}} or the Bing Autosuggest API\footnote{\url{https://azure.microsoft.com/en-us/services/cognitive-services/autosuggest/}}~\parencite{Diakopoulos-Columbia-2014}, through qualitative audits \parencite{tripodi2018searching,diakopoulos2013sex}, or by programmatically running queries, typically through dedicated browser plugins \parencite{robertson2019auditing}. 
However, due to the limited access to search logs, comprehensive studies by the broader research community are difficult to run. 
Even for those developing the search engines, the sheer number and diversity of information needs, of search log queries, as well as of possible suggestions for each given prefix that the autosuggestion mechanisms may consider surfacing, makes the efficient discovery and detection of both classes and particular instances of problematic query suggestions extremely difficult.

\vspace{4pt}
\noindent {\bf Organization \& Contributions. } 
In this paper, we provide an overview of a range of social and technical challenges when moderating web search autosuggestions.  
Our goal is to highlight difficult and on-going issues in web search autosuggestions that are still prevalent and require deeper investigation and potentially new methodologies to solve.  
We start with defining what it means for a query to be "problematic" and broadly characterizing several types of problematic query suggestions that can be surfaced, which we follow by discussing methodologies for discovering problematic queries within a large volume of queries considered by an autosuggestion system.  
We then cover technical difficulties in designing automatic methods to accurately detect and suppress problematic queries. 
We follow this with a discussion of difficult editorial decisions that must be considered when designing a suggestion moderation system.  We conclude with a brief overview of similar applications the issues we cover here extend to, and highlight a few key research directions.

\begin{figure}[t]
    \centering
    \includegraphics[width=0.70\linewidth, trim={6.4cm 6.8cm 6.2cm 4.3cm},clip]{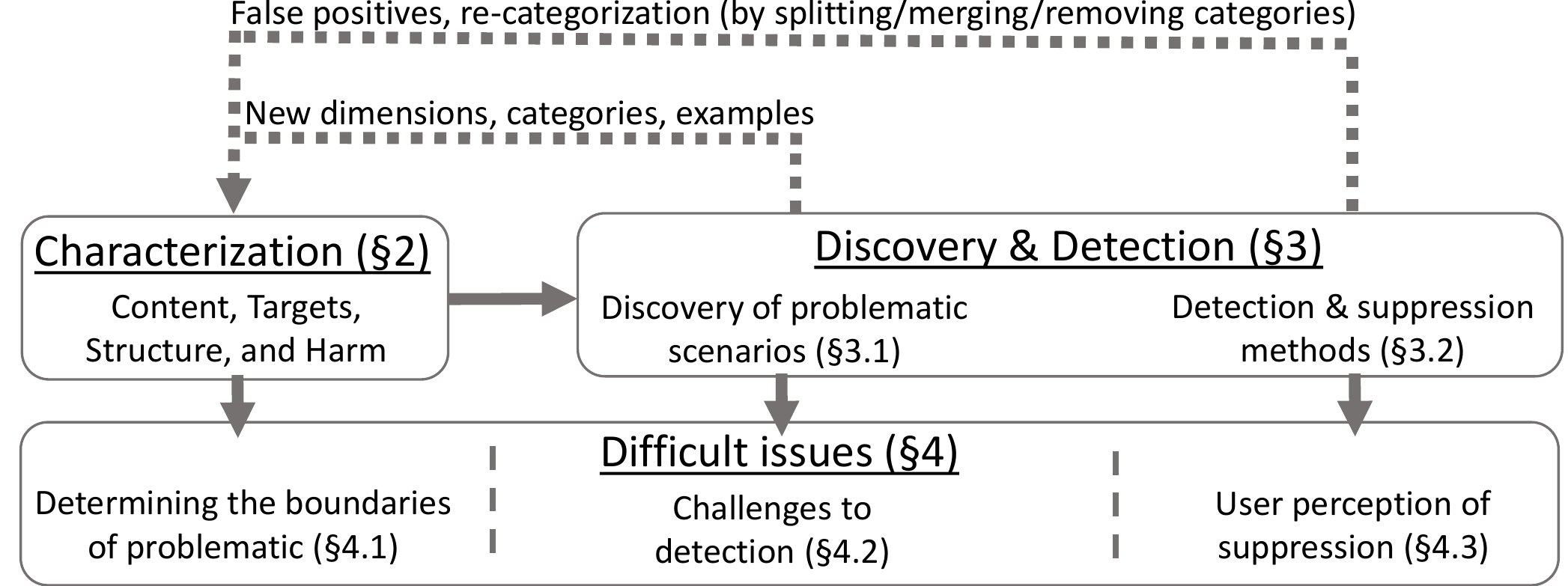}
    \caption{Organization and discussion overview.}
    \label{fig:suggestions}
\end{figure}

%% file: characterization.tex
\section{Characterization}

Identifying and mapping the universe of problematic query suggestions is difficult both due to the open domain, long-tailed nature of web search, and variations in how people appraise the potential for harms.   
In fact, there is no standard definition of what and when suggestions should be considered problematic, with existing efforts largely focusing on some well agreed upon classes of problematic suggestions such as adult content, hate and illegal speech, or misinformation.

Many of these classes, however, are themselves ambiguous with no agreed upon definition. 
Take the case of hate speech which is codified by law in many countries, yet the definitions vary widely across jurisdictions~\parencite{sellars2016defining}. 
Even for adult content there are debates about what should be included and what is harmful, and it is often challenging to make the call without proper historical or cultural context.
One such example is the iconic Napalm Girl picture, which Facebook censored and then reinstated due to public criticism \parencite{ibrahim2017facebook}.
%

\vspace{2pt}
\noindent {\bf Working definition. } 
For our discussion here---to incorporate aspects of problematic suggestions mentioned by prior work e.g., \parencite{diakopoulos2013sex,Cheung-SSRN-2015,Yenala-PAKDD-2017,Elers-Te-Kahaoa-2014}---we follow~\parencite{Olteanu-2020} and broadly consider problematic any suggestion that may be {\em unexpectedly offensive, discriminatory, or biased}, or that may {\em promote deceit, misinformation or content that is in some other way harmful (including adult, violent or suicidal content).}
Problematic suggestions may {\em reinforce stereotypes, or may nudge users towards harmful or questionable patterns of behaviour}. 
Sometimes third parties may also attempt to manipulate the suggestions to promote, for instance, a business or a website.

\subsection{Dimensions of Problematic Suggestions}

Understanding which suggestions should be construed as problematic and how to efficiently detect them also requires examining possible dimensions of problematic suggestions such as their 
%
%
{\em content} (e.g., what type of content or topics are more likely to be perceived as problematic?) \parencite{Olteanu-2020,miller2017responsible,Yenala-PAKDD-2017},
{\em targets} (e.g., who or what is more likely to be referenced in problematic queries?) \parencite{Olteanu-2020,olteanu2018effect,UN-Women-AD-Campaign-2013}, 
{\em structure} (e.g., are problematic queries likely to be written in a certain way?) \parencite{santos2017query}, and
{\em harms} (e.g., what are the harms of surfacing problematic suggestions?) \parencite{miller2017responsible}.

While these do not necessary capture all aspects that make a suggestion problematic and other relevant dimensions could and should be considered, they help illustrate the social, contextual, and topical complexity of providing web search suggestions, as well as of their consequences to users and beyond.  
They also point to the need to design and implement discovery mechanisms and frameworks (\S\ref{sec:discovery-detection}) that can help capture a wider range of scenarios.

\subsection{Content}

While existing work has mainly focused on the {\bf content} of the queries, such as recognizing profane, racist, or sexist content (e.g., \parencite{Yenala-PAKDD-2017}), other types of problematic suggestions like subtle stereotypical beliefs have often been overlooked. 
We are interested in topics and content categories which if expressed through a suggestion, then that suggestion is likely to be problematic. 
To illustrate the range of problematic suggestions, we highlight a few high-level categories that have been mentioned in prior literature, briefly discussing possible intersections and peculiarities. 
Given the long tail of problematic scenarios, however, in practice a different or more granular delimitation of the types of problematic categories might be needed~\parencite{Olteanu-2020}.

\smallskip
\noindent {\bf Harmful speech. } 
A query suggestion may constitute harmful speech if the suggested query contains 
profane language (e.g., \qs{<person>}{is a bastard}{});\footnote{To avoid disparaging any particular individual or group through our examples, we use place-holders like {\it <person>} or {\it <group>} as stand-in for any named person or group when the clarity of the query intent can be preserved.}
if it can be perceived as hateful, as 
it offends (e.g., \qs{<person>}{is dumb}{}), 
shows a derogatory attitude (e.g., \qs{girls are}{not that smart}{}),\footnote{Throughout the paper we showcase many illustrative examples for both specific types of problematic suggestions and for specific issues. While the examples have been edited for clarity and anonymity, they originate either from 1) Bing's search logs (not from existing autosuggestions), 2) an inventory of examples of autosuggestions encountered by the authors over time on various search engines, 3) prior published work, or 4) news media articles. } 
intimidates (e.g. \qs{refugees should}{not be allowed here}{}), 
promotes violence towards individuals or groups (e.g., \qs{kill all}{<group>}{}); or if it appears related to e.g.,
defamatory content promoting negative, unproven associations or statements about individuals, groups, organizations (e.g., \qs{<person>}{is a criminal}{}). 
In fact, these types of queries have been a focus of the prior work on the detection and suppression of problematic suggestions~\parencite{Gupta-ECIR-2017,diakopoulos2014algorithmic}, as they often target individuals and groups (\S\ref{sub:targets}).

\smallskip
\noindent {\bf Illicit activities and speech. }
Query suggestions may also be construed as 
illicit speech (e.g., death threats like \qs{<person>}{needs to die}{}), such as 
nudging users towards illicit activities (e.g., \qs{should I try}{heroin}{}), 
promoting terrorist/extremist content (e.g., \qs{how to join}{ISIS}{}), 
surfacing private information (e.g., \qs{<person> home address}{<address>}{}), or 
asserting information that could be perceived as defamatory (e.g., \qs{<person>}{running a scam}{})~\parencite{diakopoulos2014algorithmic,Cheung-SSRN-2015}.

\smallskip
\noindent {\bf Manipulation, misinformation, and controversy. }
Other query suggestions might be problematic if perceived as 
controversial (e.g., ~\qs{abortions should}{be paid through tax money}{}), if they seem to 
promote misinformation or information that is misleading or controversial (e.g., \qs{climate change}{is not proven}{}), or 
nudge users towards conspiracy theories (e.g., \qs{pizzagate}{is real}{}), or 
they appear manipulated in order to promote certain viewpoints or content, typically about ideological viewpoints, businesses or websites (e.g., \qs{you should invest}{in dogecoin}{}).

\smallskip
\noindent {\bf Stereotypes \& bias. }
Query suggestions might also be problematic if perceived by some users 
as discriminatory, racist (e.g., \qs{blackface is}{OK}{}), sexist (e.g. \qs{girls are}{bad at math}{,} \qs{women are bad}{managers}{}), or homophobic (e.g. \qs{gay people}{shouldn't have kids}{}), 
as validating or endorsing political views (e.g., \qs{democratic socialism}{is bad for the country}{})~\parencite{borra2012political} or ideological biases associated to certain groups (e.g., \qs{republicans are}{evil}{}), or 
as reflecting systemic biases, stereotypes or prejudice often against a group (e.g., \qs{women have}{babies for welfare}{} or \qs{immigrants}{steal jobs}{}).
Detecting biases in suggestions may also require observing systemic patterns in what is being surfaced in relation to different groups (e.g., men vs. women). 
Thus, while some of these intersect with some extreme types of harmful speech, they can also be subtle and challenging to detect.

\smallskip
\noindent {\bf Adult content. }
One of the most tackled types of problematic queries are those containing pornography related terms, or that can nudge users 
towards adult (e.g. \qs{how to find}{porn videos}{}), or obscene or racy content (e.g., \qs{naked}{women}{,} \qs{women that talk}{about <racy phrase>}{})~\parencite{diakopoulos2013sex}.

\smallskip
\noindent {\bf Other problematic categories. }
While these categories already highlight the many ways query suggestions can be problematic, other types include 
promoting animal cruelty (e.g., \qs{how to kill}{a cat}{}) \parencite{diakopoulos2013sex}, 
self-harm and suicidal content (e.g., 
\qs{ways to}{harm yourself}{}), 
reminders of traumatic events (e.g., \qs{what to do when}{someone dies}{}), or 
content on sensitive or emotionally charged topics for certain groups (e.g., \qs{what is it like}{to lose a child}{}).

\subsection{Targets}
\label{sub:targets}
Anecdotally, certain types of problematic queries are more likely to mention certain types of subjects or targets.  
In fact, looking at the prior literature (on problematic suggestions or offensive speech), a focus has been placed on content categories that often entail specific individuals or groups as the target, such as pornography (e.g., \qs{women}{naked}{}), hateful speech (e.g., \qs{arabs should}{be deported}{}) or stereotypes (e.g., \qs{men do not}{cry}{}) \parencite{Gupta-ECIR-2017,santos2017query,olteanu2018effect,davidson2017automated}. 
However, the {\bf targets} of problematic suggestions are often much more diverse~\parencite{Olteanu-2020}, including 
activities (e.g., \qs{cutting yourself is}{stupid}{}), 
animals (e.g., \qs{should i kill}{my cat}{}), 
organizations (e.g., \qs{mainstream media is}{destroying america}{,} \qs{nasa is}{a joke}{}), 
diseases (e.g., \qs{bipolar disorder is}{a fraud}{,} \qs{cancer is}{not real}), 
businesses (\qs{macy's is}{running a scam}{}), and 
religion (e.g., \qs{religion is}{stupid}), among others.

\subsection{Structural Features}

Discussed in more detail in Section \ref{sec:discovery-detection}, common templates can be identified in certain types of problematic queries, and many computational approaches to detection leverage a variety of linguistic cues. 
Indeed, it appears that {\bf structural features}, like 
how queries are formulated (syntax, terminology, or speech acts), 
how long they are, and 
how specific they are, 
might be correlated with some types of problematic suggestions \parencite{davidson2017automated,santos2017query}.
For instance, queries expanding with problematic suggestions may be structured and formulated as a sentence (declarative), question (interrogative), or as a set of terms with no grammatical structure (e.g., \qt{girls pics young}).  
They can also include different types of speech acts (e.g., assertive like \qt{young people [are crazy]} vs. expressive like \qt{upset that their [life is about to end]}), 
and can exhibit different levels of specificity (e.g., \qt{drug dealers} vs. \qt{drug dealers in new york city}) 
and varying lengths (e.g., a few terms vs. full sentences). 
Such structural properties might also correlate with the demographic attributes of those writing them \parencite{weber2010demographics,aula2005user}, and with content and target categories (\S\ref{sub:targets}).

\subsection{Harms}
\label{sub:harms}

When determining whether a suggestion is problematic, the potential for various {\bf harmful effects} 
should also be factored in (e.g., discomfort vs. physical harm).~~
Thus, beyond the legal and public relation issues caused by problematic suggestions, it has been argued that search engine companies should mitigate a wider range of potential harms. 
\citeauthor{miller2017responsible} (\citeyear{miller2017responsible}) stress that search suggestions might induce ``{\em changes in users' epistemic actions, particularly their inquiry and belief formation},'' which can have harms like ``{\em generating false, biased, or skewed beliefs about individuals or members of disempowered groups}.''
%
%
Problematic suggestions may, thus, have a variety of consequences on both individual users and the society. 


At an individual level, suggestions may nudge users towards harmful or illicit patterns of behaviors (e.g., \qs{should I try}{heroin}{,} \qs{download latest movies}{torrent}), may offend (e.g., \qs{women are} {evil}{,} \qs{immigrants are}{dirty}) or arouse memories of traumatic experience resulting in emotional or psychological harm (e.g., 
\qs{self-harm is}{selfish}{}). 
Suggestions referencing a specific individual can also harm if false or sensitive information is suggested about them, which at extremes can even constitute privacy breaches or can result in defamation.

Other types of harms include 
reinforcing stereotypical beliefs (e.g., \qs{girls are}{bad at math}{}), 
inadvertently promoting inaccurate or misleading information (e.g., \qs{bipolar disorder is}{fake}{}, \qs{vaccines are}{unavoidably unsafe}{}) or 
certain values or beliefs over others (e.g., \qs{abortion should be}{illegal}{}), 
promotion of violence towards individuals, groups, animals or in general (e.g., \qs{how to poison}{my cat}{,} \qs{how to beat}{my boss}{,} \qs{women should be}{punished}{}), or 
surfacing of controversial stances about historical events (e.g., \qs{hitler was}{right}{}), among others.  

%% file: discovery_detection.tex
\section{Discovery and Detection}
\label{sec:discovery-detection}

The implementation of a system to suppress problematic queries typically involves two key stages. 
First, there is a discovery stage where example queries are mined from search logs and identified as problematic or not (\S\ref{sub:discovery}). 
This process generates a collection of annotated examples.
Given this collection of annotated data, the second stage is to design and train machine learning (ML) models for detecting problematic queries and suppressing them from appearing as suggestions (\S\ref{sub:detection}).

\subsection{Discovery of Problematic Scenarios}
\label{sub:discovery}

A key step in mitigating problematic query suggestion---though surprisingly under-explored by prior investigations and discussions on tackling such suggestions---is their {\em discovery}.
Though current methods for discovering problematic queries are varied, they are often ad-hoc in nature, usually requiring a combination of human intuition (typically of system designers) with automated detection methods. 
This can leave them prone to important blind spots, which we overview below.

\subsubsection{User Reporting}
Search engines provide mechanisms for users to report problematic suggestions.\footnote{The link to report offensive autocomplete predictions on Google is:  \url{https://support.google.com/websearch/contact/report_autocomplete}. 
For Bing, feedback on offensive suggestions can be submitted by clicking the "Feedback" link on the bottom ribbon of the Bing page containing the problematic suggestion. } 
User reporting is invaluable to surfacing problematic suggestions missed by automatic detection mechanisms, and typically results in the timely removal of the specific suggestions being reported. 
Such fixes, however, are reactive in nature and may be skewed towards the most salient cases, overlooking lesser known, cultural or context sensitive scenarios.

\subsubsection{Red Teaming}
While user reporting is helpful, search engine providers prefer to discover problematic queries proactively and remove them before they are observed by any users. For this, search engine providers may employ ``red teams,''\footnote{\url{https://en.wikipedia.org/wiki/Red_team}} i.e., teams of independent workers tasked with probing a search engine for problematic suggestions. 
Red teaming seeks to expose weaknesses or blind spots in the search engine's detection mechanism or even new classes of problematic suggestions that the system developers were unaware of. 
This process can mimic white hat or ethical hacking practices~\parencite{caldwell2011ethical,palmer2001ethical},\footnote{\url{https://en.wikipedia.org/wiki/White\_hat\_\%28computer\_security\%29}} where individuals can specialize in identifying such ``vulnerabilities.''
	

\subsubsection{Exploring Common Templates}
After an initial discovery phase from user reports or red teaming exercises, sets of problematic queries can be explored to find common templates that frequently carry derogatory suggestions. Template based approaches that leverage the syntactic structure of queries when collecting and sampling data, can for instance help identify hateful and offensive speech with higher precision; like \qt{I <intensity><user intent><hate target>} to identify targets of hate speech (e.g., \qt{I really hate educated women})\parencite{davidson2017automated,gitari2015lexicon,silva2016analyzing}. Even simple query prefixes like \qs{is <person>}{...}{} or \qs{<person> is}{...}{} are often completed with derogatory phrases. By collapsing queries into these common templates and then annotating the high frequency templates, additional derogatory phrases used to describe people can be discovered quickly. 

%

\subsubsection{Query Embedding}

To help discover novel query forms expressing the same semantics as known problematic queries, deep-learned semantic embeddings can be used. Query embeddings can be learned from web search click data to create a vector embedding space in which queries with similar click patterns are placed close together in the embedding space~\parencite{Huang-CIKM-2013,Shen-WWW-2014}. By exploring queries located close to known problematic queries in an embedding space, new problematic query forms and derogatory phrases can be discovered. 
However, this is sensitive to the seed set of known problematic queries, and is likely to miss cases that differ in nature. 	
	

\subsubsection{Active Learning}

To take advantage of both machine learning models and human effort, an active learning approach can be taken. Active learning is a human-in-the-loop method where machine learned detection models (such as those we discuss later in Section~\ref{sec:ml_models}) can examine large volumes of unannotated queries and propose the queries most likely to improve the model if annotated~\parencite{Settles-2009}. Typically these are borderline queries in the model's score space that the model has uncertainty on. By iteratively repeating the annotation of new queries proposed by the model and then retraining the model with the new data, both new problematic queries can be discovered and the ML model can be improved.

\subsubsection{Linkage to Problematic Web Sites}

The search results returned for a given query can also be used to assess a query's propensity to be problematic. 
If the returned results are in some way problematic (e.g., contain adult content, low quality), the query itself may also be problematic~\parencite{parikh2012identifying}.
Independent of the actual content of a query, a search engine should not suggest any query that if selected would return as search results problematic web sites. 
This could include sites whose content is pornographic, incites violence, promotes illegal activity, or contains malware. 
Because of the potential for black hat adversaries to manipulate search results for particular queries (including those of a seemingly benign nature), discovering such queries may require the detection of problematic web sites in the search results returned for the query~\parencite{Lee-IEEE-2002,Arentz-CVIU-2004,Wang-DSN-2013}.

\subsection{Detection \& Suppression Methods}
\label{sub:detection}

Because of the complexity of the problem and the potential harm of surfacing problematic queries, search engines typically employ a mix of multiple
manual and automated methods for detecting and suppressing problematic queries in order to improve robustness~\parencite{santos2017query}. Some of the most common approaches are discussed below.

\subsubsection{Block Lists}

It is common for systems to maintain manually curated block lists of highly offensive terms that will trigger the suppression of a suggestion in any circumstance. 
However, many terms are offensive only in certain contexts, e.g., {\it scum} is an offensive suggestion in \qt{<person> is [scum]} but non-offensive in \qt{what is pond [scum]}. Therefore, detection techniques need to model the entire query to avoid falsely suppressing legitimate query suggestions.

Block lists for whole queries are also employed and can be implemented efficiently for use in a run-time system. These lists allow for on-the-fly updating if a content moderator needs to immediately suppress an offensive query suggestion reported by a user. Block lists also ensure previously flagged queries remain permanently suppressed even if other modeling techniques unexpectedly leak the query after a model update. However, block lists cannot be generalized easily and are not a scalable solution by themselves.

\subsubsection{Query Templates and Grammars}

Because search queries tend to be short and the likelihood of query uniqueness increases with query length, the collection of common query candidates used by an autosuggest feature will be dominated by short queries.\footnote{The search queries were found to have an average length of about 4-5 words~\parencite{robertson2019auditing}.} As such it becomes possible to cover the head of the distribution of derogatory queries with simple templates or finite state grammars. 

For example, to suppress derogatory suggestions about named individuals a simple approach is to use an entity extractor to identify people's names within queries, curate a list of derogatory expressions, and then identify the most common query templates that use a combination of a person's names and a derogatory expression such as
\begin{inparaenum}[]
    \item \qt{<person> is <derogatory\_expression>}, 
    \item \qt{is <person> <derogatory\_expression>}, or
    \item \qt{why is <person> so <derogatory\_expression>}.
\end{inparaenum}

To cover the wide range of template variations, a more efficient approach is to encode them into a rule-based grammar using a regular expression based finite-state tool (e.g., Foma~\parencite{Hulden-EACL-2009}). This has the advantage of quickly generalizing the head of the derogatory query distribution with high precision, but can suffer from poor coverage for longer or less-common query forms. It also requires a human to update the grammar which is not scalable for covering tail queries and improving recall.

\subsubsection{Machine Learning Models}\label{sec:ml_models}

To achieve greater generalization and avoid the use of hand-written rules, ML approaches can also be used to detect problematic queries. Techniques that have been explored for this purpose include gradient boosting decision trees~\parencite{Chuklin-Sexi-2013}, long short-term memory networks~\parencite{Yenala-PAKDD-2017}, and the deep structured semantic model~\parencite{Gupta-ECIR-2017}. Relative to grammar-based models, ML models have been found useful in improving the detection rate of problematic queries but at the expense of increased false detections~\parencite{Gupta-ECIR-2017}. While ML models avoid the effort required to hand-craft rules, they do require human effort to annotate collections of queries (both problematic and non-problematic) in order to train a model.

\subsubsection{N-strike Rule}

One scenario that can plague a query suppression mechanism occurs when a system successfully removes problematic queries from a suggestion list only to have them replaced 
by other problematic suggestions missed by the detection model. This situation is even more likely when adversarial attacks specifically seek to find holes in the model. One way to combat this is to employ an $N$-strike rule, i.e., the detection of $N$ or more problematic queries at the top of a pre-filtered suggestion list for a query prefix will suppress all suggestions for that prefix. 

While it may be unclear for any single, arbitrary example what mechanisms were used to suppress suggestions, the complete suppression of suggestions for certain problematic prefixes is observable in both Google and Bing. For example, neither search engine provides any suggestions for prefixes such as \qt{jews are,} \qt{muslims are,} and \qt{catholics are,} which, without such interventions, did yield multiple highly inappropriate suggestions in the past \parencite{Guardian-2016}. 


%% file: challenges.tex
\section{Difficult Issues}

There are lingering challenges to addressing these issues that require additional research, maybe even re-thinking current approaches to issues of  characterization, detection, and suppression; challenges that we cover next.   



\subsection{Setting the Boundaries of "Problematic"}

\subsubsection{Operationalizing ``Problematic.''}
Even with a clear definition of what constitutes ``problematic'' query suggestions, setting the boundary between problematic and non-problematic cases can still be difficult. 
For example, certain system provided suggestions may still be deemed offensive by some people but not by others.  
The typical way to determine whether suggestions are problematic is through some form of crowd labelling or block lists (for both data collection and annotation), but both  approaches have limitations when faced with more ambiguous, latent concepts. 
For instance, there are many factors that affect human assessments including ambiguous definitions, poor annotation guidelines, poor category design and insufficient context~\parencite{olteanu2019social}. 
For concepts like hate speech, user or crowd judge characteristics may also lead to variations in how different users or judges make assessments even when given the same definition~\parencite{olteanu2017limits}.

\subsubsection{``Problematic'' in Context}
Another issue is that the context in which a suggestion is surfaced can also impact how that suggestion is perceived. 
In fact, there are cases where the suggestion is problematic because of the context (or the lack thereof). 
Consider the examples: 
\qs{why do teens}{commit suicide}{,}  
\qs{can you purchase}{a gun online}{,} or
\qs{what does the}{bible say about abortion}{.}
Such suggestions might be problematic since what the user wrote (the query prefix) did not warrant the surfacing of sensitive or controversial topics. 
Getting the context right in these cases is often hard. 

In addition, how various types of problematic suggestions are expressed can vary across contexts and over time. 
In addition to active forms of obfuscation to trick the engine detection and suppression components (\S\ref{sub:detection_issues}), the way in which, for instance, harmful speech is expressed is not always static and predictable, but it varies 
based on the identity of the author and of the target, 
based on the time of the day or 
based on some world events~\parencite{schmidt2017survey,cheng2017anyone,kumar2017antisocial}, among others.

\subsubsection{Subjectivity}

Personal beliefs about socially divisive topics such as religious beliefs, gender identity, abortion, and immigration, also contribute to the subjective perceptions of what is {\it problematic}. For example, a query suggestion of the form \qs{<person> is}{gay}{} may be viewed differently by people depending on their own personal views and who the target of the query is. 
For someone who proudly self-identifies as being gay, being the target of this suggestion is likely not problematic. However, someone with strong anti-LGBTQ beliefs may perceive this suggestion as highly offensive, especially if they are the target. 
In general, the perceived offensiveness of such a query can vary across the general population as the term {\it gay} is commonly used in a non-offensive manner to convey sexual identity~\parencite{GLAAD-Guidelines}, but is also widely used as a pejorative~\parencite{Winterman-BBC-2008}. 
Consideration should be given to how assessments of what is problematic should be handled, and whether e.g., a `majority' agreement (often the standard for crowdsourcing assessments) is an appropriate criteria for flagging suggestions as problematic.  

%


\subsubsection{Truthfulness \& Mislabeling}
A key driver for suppressing suggestions is preventing defamatory statements from appearing as suggestions. For example, a query form \qs{serial killer}{<person>}{} could be viewed as defamatory if the person is not a serial killer.
If the person is verifiably a serial killer, then the query is not defamatory and could be a common legitimate query that need not be suppressed. Avoiding suppression in such cases should require verification from a trusted knowledge base 
before showing the query.

A pernicious form of untruthfulness is intentional mislabeling. As described in~\parencite{Molek-Kozakowska-2010}, intentional mislabeling is a political ploy for "{\em introducing and/or propagating terminology that is either inaccurate or derogatory (or both) to refer to a person, a group, or a policy in order to gain political advantage.}" Such mislabeling is often used to imply someone is different from what they proclaim to be and/or suggest a difference that is perceived negatively by the intended audience of the messaging. A high profile instance is the mislabeling of Barack Obama as a Muslim by his political opponents~\parencite{Holan-Politifact-2010}.  

Mislabeling is particularly difficult to detect in autosuggestions because the terminology used is often not problematic in its own right, and would go undetected by typical filtering methods. Other mislabeling forms involve political stance (e.g., \qs{<person> is a}{communist}{}) and gender (e.g., \qs{<person> is a}{man}{}).

\subsubsection{Newsworthiness}

In some situations, clearly offensive query suggestions result directly from current newsworthy events, particularly when prominent individuals make derogatory statements about other people. 
In news media, editorial decisions are made to weigh the potential harm of reporting offensive content against the public's need for full knowledge of an event. 
A recent example that news editorial teams struggled with was Donald Trump's profanity laden reference to African countries~\parencite{Jensen-NPR-2018}.
Similarly, if an autosuggestion mechanism has the ability to discern when a query is referencing a current news story, it may also consider temporarily allowing a problematic query as an autosuggestion while the query's subject matter is active in the news cycle. 
This approach would provide easier access to legitimate news stories while they are active, but suppress them when they fall out of the news cycle. 


\subsubsection{Historical Inquiry}


In general, negative queries about long deceased persons are unlikely to be submitted for the intent of defaming the person and are more likely submitted for the purpose of historical inquiry.
%
In cases where a historical figure is involved (e.g., \qt{was woodrow wilson [a racist]}), suppression of such query suggestions may be unnecessary. Again, allowing such  suggestions would typically require verification that the target is a deceased person of historical importance from a trusted knowledge source.

\subsection{Challenges for Detection Methods}

\label{sub:detection_issues}

Even with a good understanding of what constitutes problematic scenarios, detecting abusive or other types of problematic language can be difficult. In some cases, they can be subtle, fluent, formal, and grammatically correct, while in others they can be ambiguous and colloquial ~\parencite{nobata2016abusive}. 
Other challenges involve adversarial queries, data voids, and others; which we discuss next. 

\subsubsection{Adversarial Queries}

As in any adversarial situation, people intending to corrupt autosuggestions for their own purposes will probe for novel ways to defeat any algorithmic suppression model. Adversaries may use a variety of tricks to mask the intent of a query from automated detection while still providing a clear intent to a typical user. 
A common circumvention strategy is to rewrite an offensive or problematic query to include misspellings, abbreviations, acronyms, homophones, leet speak, or other text manipulations. 

Prior work particularly focused on hate and offensive speech had documented such strategies that circumvent abuse policies and automated detection tools.
To avoid being detected by the automated hate-speech detection tools, users have developed a code (e.g., the operation Google Movement) in which references to targeted communities are substituted by ``benign'' terms in order to seem out of context~\parencite{magu2017detecting}. 
Similarly, subtle changes in the toxic phrases 
(e.g.  ``st.upid'', 
``idiiot'') 
have also proved effective in deceiving automated tools like Google's Perspective, while also being susceptible to false positives (e.g., by not correctly interpreting negations)~\parencite{hosseini2017deceiving}.

Examining search logs from Bing, we found similar examples including  
\begin{inparaenum}[]
\item \qt{<person> is an eediot,} 
\item \qt{<person> is a a hole,} 
\item \qt{<person> is a pos,} or
\item \qt{<person> is a knee gar.}
\end{inparaenum}
Indeed, misspelled words---which are also less competitive in terms of the traffic they draw---have become a target for manipulation, with adversarial parties employing increasingly sophisticated techniques to circumvent counter-measures that e.g., do automated corrections~\parencite{joslinmeasuring}.

\subsubsection{Data Voids}

The open ended nature of search allows users to look up anything and everything. 
Yet, some topics and their corresponding queries will be more popular than others---more users will search for \qt{treating a cold} than for \qt{quantum computing} or \qt{asexuality}.
This leads to topics and associated queries for which there is little to {\em no} 
web content.
%
Instances where ``{the available relevant data is limited, non-existent, or deeply problematic}'' have been dubbed as {\em data voids}~\parencite{golebiewskid}.
For instance, the anti-vaccine movement is believed to have leveraged existing information voids to promote their beliefs within the top results~\parencite{wired2019}.
Similarly, there will also be queries (and query prefixes) that are less frequent, making suggestions much more prone to manipulation and errors---e.g., such as misspelling of popular queries~\parencite{joslinmeasuring}. 
One such example in Bing's search logs is \qt{<health insurance> of ill}, where ``ill'' is a common misspelling of the Illinois state's abbreviation (IL).

In fact, each query tends to be associated with its own unique demographic fingerprint \parencite{shokouhi2013learning}, with rare queries being frequently run by niche groups of users, some of which can be adversarial in nature. 
Leveraging this at run-time can be computationally intensive, and may raise privacy and bias concerns (if e.g., queries from certain groups are more routinely suppressed).

\subsubsection{Derogatory Queries and Trolling}

Prominent public figures, especially polarizing politicians or partisan political commentators, are particularly susceptible to the appearance of derogatory autosuggestions culled from user queries. While ``trolling'' of such individuals likely happens, it is often unclear whether any particular derogatory query suggestion is the result of a coordinated ``trolling'' attack or simply arises through organic searches submitted by users with negative views of an individual. For search engines that frequently update their autosuggestion query candidates, negative news events can generate large query frequency spikes that could introduce new derogatory suggestions about such individuals.

\begin{figure}[t]
  \begin{center}
  \includegraphics[width=1.0\textwidth]{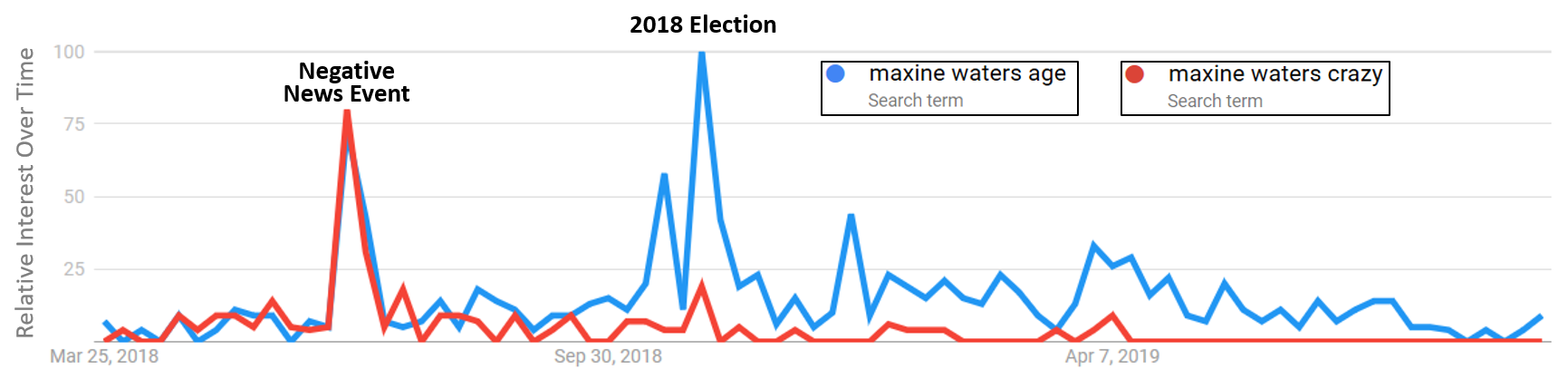}
  \includegraphics[width=1.0\textwidth]{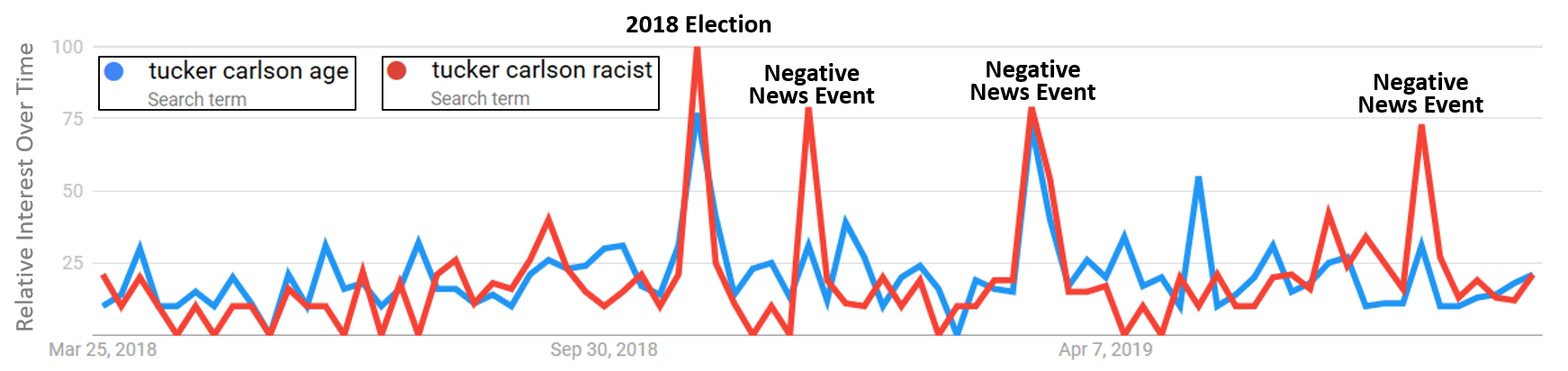}
  \end{center}
  \caption{Relative frequency of submissions to Google's search engine of a neutral fact based query (in blue) and a derogatory query (in red) about Maxine Waters in the top graph and Tucker Carlson in the lower graph over an 18 month period spanning March 2018 through September 2019. The relative query frequencies were generated using the Google Trends online tool.}
  \label{fig:Waters_Carlson}
\end{figure}

Figure~\ref{fig:Waters_Carlson} shows query trends for two prominent and politically polarizing figures in the United States. Each plot shows the relative query volume for a neutral fact based query (in blue) versus a derogatory query (in red) over an 18-month time period. By examining the query volume for each individual, it is easy to find spikes in volume following events that puts them in the public eye. Further, the volume of negative queries typically spike and often exceed the volume of neutral queries when news events paint them in a negative light. In systems whose suggestions are ranked using query frequencies within a recent time window, highly frequent negative queries can replace more neutral queries in the suggestion list and can remain there while their observed frequency spikes remain within the analysis time window. Rapid changes in query volume about individuals can be cues for moderators to examine prefixes associated with them for potential trolling. 

\subsubsection{Cultural Sensitivities}

The offensiveness of queries may be limited to certain national, ethnic or cultural sub-populations. Without an awareness of the derogatory terminology or social sensitivities particular to a sub-population, it can be difficult for a system's developers or annotators to recognize some queries as problematic. 



To provide an example, the phrase "black and tan" in the United States refers to a layered beer cocktail. 
However, in Ireland it refers to the military forces sent to Northern Ireland to suppress the Irish independence movement. 
Because of the brutal tactics employed by the group, the phrase is primarily pejorative in Ireland~\parencite{Bell-Vinepair-2016}.
In designing models and policies for content moderation it is prudent to avoid a lack of diversity across different ethnic, gender, political, or cultural groups in the pool of people creating, maintaining, and governing such content moderation systems.

\subsubsection{Stereotyping} 
Common ML algorithms learn features and frequent associations about words and phrases, but they are often incapable of capturing deeper insights into the social or cultural aspects of a query. 
As such, it may be difficult for them to make nuanced decisions about queries that reinforce adverse stereotypes. For example, consider the suggestion \qs{girls should play}{softball}{} versus the suggestion \qs{girls should play}{baseball}{.} 
There is nothing inherently problematic about girls playing either softball or baseball, but the former suggestion can subtly reinforce the stereotype that girls should not play baseball (a sport historically played primarily by boys) while the latter can be an empowering statement to enable girls to disregard the stereotype.  
Research into discovering common gender stereotypes present in word embedding vectors examines this issue~\parencite{Bolukbasi-NIPS-2016}, but these solutions are sensitive to various implementation parameters, and solving this problem for the wide range of social and cultural stereotypes possible in autosuggestions remains an open problem.

\subsubsection{False Suppression}

It is sometimes difficult to distinguish a derogatory query from a legitimate non-offensive query based only on the query's words. 
Some non-offensive queries can appear offensive when viewed without deeper knowledge of the intent. Consider these example suggestions: \begin{inparaenum}[]
\item \qs{stupid girl}{jennifer nettles}{,}
\item \qs{judd apatow}{sick in the head}{,}
\item \qs{prince william}{trash}{.}
\end{inparaenum}
Without additional context these suggestions may appear to be derogatory statements about the mentioned individuals. Yet, "Stupid Girl" is actually a song by Jennifer Nettles, "Sick in the Head" is a book by Judd Apatow, and the query \qt{prince william trash} likely references a trash disposal service in Prince William County, Virginia and not Prince William the individual. Such examples show that only examining a query surface form is sub-optimal and can result in legitimate queries being suppressed. 

Additional mechanisms that take advantage of knowledge gleaned from prior search results or other query understanding models can help mitigate such false suppressions.  For example, the non-offensive nature of the queries above can be identified through the use of web search entity linking and detection tools that match these queries against entities present in a knowledge base (e.g., titles of works of art, business names).  
Of course, references to works of art containing highly offensive or profane phrases in their titles can still be suppressed through the use of phrase block lists.

More broadly, dealing with ambiguity has been found challenging in other similar settings. 
For instance, sarcasm can be confused with abusive language~\parencite{nobata2016abusive}, while ambiguity may originate from both the use of language and how various types of problematic suggestions are defined. 

\subsubsection{Promotional Manipulation}

Evidence exists that a cottage industry of black hat search engine optimization (SEO) businesses focused on manipulation of search engines' autosuggestion feature has emerged. Their goal is to embed queries promoting businesses or products into search autosuggestion lists for key query prefixes \parencite{Vernon-2015}. These suggestions can be particularly problematic if the activities of the promoted business are illicit (e.g., promoting software installations containing malware). These manipulations can be difficult to detect based on the query alone as they are designed to resemble similar legitimate queries, though there are common search templates within which these manipulations typically occur. One study  estimated that roughly 0.5\% of query suggestions covering 14 million common query prefixes exhibit evidence of SEO manipulation. These manipulations covered a range of business sectors including home services, education, legal and lending products, technology, and gambling~\parencite{Wang-NDSS-2018}.

\subsection{User Perceptions of Suppression}

\subsubsection{Is Suppressing Suggestions Censorship?}

The suppression of query suggestions has been sometimes portrayed as a form a censorship in the press and in online blogs~\parencite{Anderson-Hobo-2016,Wilson-Betanews-2016}. This argument assumes that user queries themselves represent a form of content and search engines that return autosuggestions are providing a platform for exposing common user viewpoints via the queries the users submit. The removal of derogatory suggestions about individuals has been particularly derided by some political figures for its potential to direct users away from negative information about their opponents~\parencite{Akhtar-USAToday-2016}. Search engines defend their practices by indicating that the suppression of suggestions does not prevent users from submitting search queries of their choosing, nor does it alter the content that is returned for any such search query~\parencite{Yehoshua-Blog-2016}.

\begin{table}[t]
  \begin{center}
    \footnotesize
    \begin{tabular}{|l|l|} 
      \hline
      \textbf{Before Suppression}   & \textbf{After Suppression} \\
      \hline
      <person> is \bf{a liar}  & <person> is \bf{back} \\
      <person> is \bf{a joke}  & <person> is \bf{right} \\
      <person> is \bf{evil}    & <person> is \bf{awesome}  \\
      <person> is \bf{a moron} & <person> is \bf{king} \\
      <person> is \bf{wrong}   & <person> is \bf{great} \\
      <person> is \bf{dumb}    & <person> is \bf{from where} \\
      \hline
    \end{tabular}
  \end{center}
  \caption{Example of potential effects of suppression of derogatory queries for the top of an autosuggestion list for the prefix form "<person> is" both before and after the query suppression model is applied.}
  \label{tab:before_after}
\end{table}

\subsubsection{Avoiding Bias}

There is evidence that the general public feels that major technology companies are politically biased and intentionally suppress certain viewpoints~\parencite{Tiku-Wired-2018}. 
Some have also suggested that suppressions within search autosuggestions are purposely favoring one political party over another~\parencite{Project-Veritas-2019}. Search engines defend their query filtering mechanisms by claiming they apply the same filtering policies regardless of the political affiliations of the individuals or groups mentioned in the queries~\parencite{Akhtar-USAToday-2016}. 

That said, major search engines serve a wide range of users and should thus strive to avoid even the perception of bias in their content moderation. Because humans may innately possess unconscious bias against certain points of view or people, it is important to remove such biases from the process of annotating data and designing models. For example, a person's political views may influence their perception of a query suggestion. Consider a suggested query of the form \qs{<person>}{is corrupt}{;} this query template would generally be considered as a derogatory and potentially defamatory statement about the person. 
However, if we show this query form using a particular person's name in place of the {\it <person>} marker, an annotator's perception of the offensiveness of the query may be reduced or heightened based on their personal opinion of the mentioned person. Designers of annotation tasks should remove or account for the potential for such biases wherever possible.

Additionally, biases from a search engine's user population can also seep into the moderation process \parencite{olteanu2019social}. This can, for instance, occur through the feedback mechanisms that allow users to report content or suggestions as offensive. 
If a search engine's user base is e.g., skewed in favor of a particular political viewpoint, then suggestions deemed derogatory about politicians they agree with may be reported, reviewed by a moderator, and subsequently suppressed more frequently than suggestions about politicians with an opposing viewpoint. It is thus important for the moderation process to not only suppress the unique queries reported by users in an online block list (which is a common for immediate suppression of offensive content) but also to update the detection model so the offensive query form is suppressed for other individuals as well.

\subsubsection{Introduction of Positive Bias}

Even if care is taken to avoid bias during the moderation process, the presence of query suppression can become obvious to users and cause them to perceive a suggestion list as biased or intentionally manipulated. 
For example, when applying an approach where derogatory, defamatory or offensive queries are suppressed, the suggestion list can become susceptible to a reverse form of {\it positive bias}. For a highly polarizing public figure, the list of queries submitted about that person will contain a mix of both highly positive and highly negative queries, in addition to neutral fact-oriented queries. 
In adversarial settings, their autosuggest list without any suppression would likely be dominated by highly negative queries like those in the "Before" column in Table~\ref{tab:before_after}. However, after removing the derogatory queries, their autosuggest list might end up looking abnormally positive in nature, such as in the "After" column in Table~\ref{tab:before_after}. While the positive queries that remain after suppression are not offensive in nature, they may appear abnormally biased in the positive direction given the polarizing nature of an individual. In such cases, a neutral approach where only queries that don't express either a positive or negative sentiment could help avoid the appearance of bias. 


%% file: discussion.tex
\section{Applications \& Future Directions}

We identified a variety of lingering issues with web search autosuggestions, which point to research areas that require renewed attention. 
Among these, critical areas include tackling biases and subjectivity, understanding the impact of moderation related interventions (or their lack off), and distilling the role of contextual cues~(\S\ref{sub:key_topics}). 
The issues we described are more and more pressing as they apply to a growing number of applications (\S\ref{sub:applications}).

\subsection{Key Research Topics}
\label{sub:key_topics}

\noindent {\em The issues surrounding bias \& subjectivity are particularly difficult. }
There are instances where suggestions would not be deemed problematic by a majority of users, but could (justly) be viewed as highly problematic by a (small) minority. 
This highlights the difficulties search engines face when defining and executing a policy, especially when judgements on whether queries are problematic are crowd-sourced (as is often the case). 
Where should the line be drawn on problematic queries and how it affects users' experience?

To avoid the appearance of bias, search engines should, as much as possible, avoid promoting specific opinions within their suggestions in favor of providing neutral or objective suggestions. 
Yet, even when all candidate suggestions are neutral, bias could still creep in and result in certain types of associations or references being surfaced in certain contexts but not others. 
How should candidate suggestions be mixed and varied across users, time, and contexts to avoid either obfuscating or promoting certain associations?

\noindent {\em It is unclear the extent to which moderation of suggestions affects search engine utility or user satisfaction. } 
Would users find suggestions more helpful if they were unbiased and focused on neutral information seeking? 
Or, do users prefer to see suggested queries submitted by other users even when they are biased, offensive or problematic in other ways? 
How should the impact of moderation-related interventions (or their lack thereof) be appraised?
When suggestions are both useful (matching the user's search intent) and problematic, should a search engine still surface them?\footnote{E.g., if a user believes that garlic cures cancer, should suggestions suggest that this is in fact true: \qs{garlic}{cures cancer}{}?} These are all open questions that require additional study to understand the actual and perceived effects of suggestion moderation on search.

\noindent {\em With misinformation rampant on the internet, a major issue is whether suggestions contribute to its spread.} 
Many suggestions may not appear problematic on their own but can be problematic if they help promote misinformation, such as directing users toward unfounded conspiracy theories, or do harm in other manners. 
To combat this requires deeper investigations into the sites that suggested queries point to and their authority.

\noindent {\em Some contextual factors make search engines more prone to problematic suggestions. }
They can also affect how users perceived certain suggestions. 
Yet, it is often hard to both know which contextual factors to consider and to distill their impact. 
Frequent queries are more likely to be suggested, but we may not know why they are frequent: do they reflect some prevalent needs or attempts to manipulate the search engine?


\subsection{Extensions to Other Applications}
\label{sub:applications}

Search query suggestions are part of a broader class of applications that provide {\em text prediction} or {\em text rewriting} suggestions to help users write faster, better, or in a more inclusive manner~\parencite{arnold2018sentiment,kannan2016smart,Cai-FTIR-2016}.
%
%
These applications aim to reduce the writing burden and help users efficiently complete tasks. 
However, they often face similar challenges to issue discovery and detection, as they operate in a similar fashion as the search autosuggestion engines do: 
{\em using past writing samples (or usage logs) and limited context, they need to contribute suggestions for an open domain, long-tailed set of requests and needs}. 
Such applications span web search, email and chat response and composition, as well as more general purpose assistive writing.

\noindent {\em Other Search Applications.}
Besides providing completion suggestions as users type their queries, search engines also offer similar query suggestions via features like ``{\em Related Searches},'' ``{\em Top Stories},'' or ``{\em People also searched for}.''~
While these features differ from search autosuggestions, as full queries or even topics are being suggested, their appropriateness similarly depends on factors like cultural or historical context, or newsworthiness. 
As a result, they are similarly sensitive to societal biases, data voids and manipulation.

\noindent {\em Email and Chat Responses. } 
The use of predictive text is also increasingly common in email and other conversational environments. 
Relevant applications include the now popular Smart Reply \parencite{kannan2016smart}
and Smart Compose~\parencite{chen2019gmail} features, which provide textual suggestions to users in the form of brief, standalone messages 
or of partial sentence completions. 
While the systems generating these suggestions often leverage richer and cleaner data, they were also found to surface problematic suggestions that e.g., misgendered users~\parencite{theverge2018}, or surfaced offensive associations \parencite{cnn2017}.
Due to their more personal use nature, these applications have also raised issues concerning the possibility of impersonating users~\parencite{gupta2018impersonation}.

\noindent {\em General Purpose Assistive Writing. }
More broadly, predictive text suggestions are used to assist users in any writing environment \parencite{arnold2016suggesting,hagiwara2019teaspn}, including by making re-writing suggestions that make writing more concise and inclusive~\parencite{bbc2019}.
Yet, due to the linguistic and social complexities of natural language that also affect the search query suggestions, such assistive writing systems can and have similarly failed to account for e.g., socio-cultural sensitivities~\parencite{medium2019}.

\section{Conclusions}

We have highlighted some of the social and technical challenges that make the moderation of web search autosuggestions difficult. While great progress has been made in suppressing the most obvious and harmful problematic suggestions, problems still remain as the query space has a long tail.  
In addition, the implementation of similar features across other applications may raise new issues. 
We hope some of the most challenging aspects of the problem will be addressed with new research and development efforts. 

Greater transparency about the issues can help alleviate some of the concerns about the practices of autosuggestion moderation. For example, Google has published a policy which provides a high level description of the types of queries that it suppresses.\footnote{Google's policy is posted at: \url{https://support.google.com/websearch/answer/7368877}} Sustained efforts by search engine companies to improve the processes and deployed technologies will also help minimize harms and increase public trust in this generally helpful search feature.

%% file: Autosuggest - arxiv/Autosuggest.bib
@online{Akhtar-USAToday-2016,
  author       = {A. Akhtar},
  howpublished = {USA Today},
  title        = {{Google defends its search engine against charges it favors Clinton}},
  url          = {https://www.usatoday.com/story/tech/news/2016/06/10/google-says-search-isnt-biased-toward-hillary-clinton/85725014/},
  year         = {2016},
}

@online{Anderson-Hobo-2016,
  author       = {S. Anderson},
  howpublished = {Hobo, UK SEO Services},
  title        = {{Factual SEO: Is Google Censoring Negative Searches about Hillary Clinton?}},
  url          = {https://www.hobo-web.co.uk/censoring-autocomplete-results-for-celebrities/},
  year         = {2016},
}

@Article{Arentz-CVIU-2004,
  author  = {Arentz, W. A. and Olstad, B.},
  journal = {Computer Vision and Image Understanding},
  title   = {Classifying offensive sites based on image content},
  url     = {https://www.sciencedirect.com/science/article/abs/pii/S1077314203001875},
  year    = {2004},
  volume  = {94},
}

@InProceedings{arnold2018sentiment,
  author       = {Arnold, K. and Chauncey, K. and Gajos, K.},
  booktitle    = {Proc. of Graphics Interface},
  title        = {Sentiment Bias in Predictive Text Recommendations Results in Biased Writing},
  url          = {https://www.eecs.harvard.edu/~kgajos/papers/2018/arnold18sentiment.pdf},
  year         = {2018},
}

@InProceedings{arnold2016suggesting,
  author    = {Arnold, K. and Gajos, K. and Kalai, A.},
  booktitle = {Proc. of ACM Symposium on User Interface Software and Technology (UIST)},
  title     = {On Suggesting Phrases vs. Predicting Words for Mobile Text Composition},
  url       = {https://dl.acm.org/doi/pdf/10.1145/2984511.2984584},
  year      = {2016},
}

@Article{aula2005user,
  author    = {Aula, A.},
  journal   = {Universal Access in the Information Society},
  title     = {User study on older adults' use of the Web and search engines},
  year      = {2005},
  number    = {1},
  volume    = {4},
  publisher = {Springer},
  url       = {https://link.springer.com/content/pdf/10.1007/s10209-004-0097-7.pdf},
}

@Online{bbc2019,
  author       = {{BBC}},
  title        = {{Microsoft Word AI `to improve writing'}},
  howpublished = {BBC News},
  url          = {https://www.bbc.com/news/technology-48185607}, 
  year         = {2019},
}

@Online{Bell-Vinepair-2016,
  author       = {E. Bell},
  howpublished = {Vinepair},
  title        = {Why You Should Never Order A Black And Tan In Ireland},
  url          = {https://vinepair.com/wine-blog/why-you-should-never-order-a-black-and-tan-in-ireland},
  year         = {2016},
}

@InProceedings{Bolukbasi-NIPS-2016,
  author    = {Bolukbasi, T. and Chang, K. and Zou, J. Y and Saligrama, V. and Kalai, Adam T},
  booktitle = {Proc. of Conference on Neural Information Processing Systems},
  title     = {Man is to Computer Programmer as Woman is to Homemaker? Debiasing Word Embeddings},
  url       = {https://papers.nips.cc/paper/6228-man-is-to-computer-programmer-as-woman-is-to-homemaker-debiasing-word-embeddings.pdf},
  year      = {2016},
}

@Article{borra2012political,
  author  = {Borra, E. and Weber, I.},
  journal = {First Monday},
  title   = {Political Insights: Exploring partisanship in Web search queries},
  year    = {2012},
  number  = {7},
  volume  = {17},
  url     = {https://firstmonday.org/ojs/index.php/fm/article/view/4070/3272},
}

@Article{Cai-FTIR-2016,
  author  = {Cai, F. and de Rijke, M.},
  journal = {Foundations and Trends in Information Retrieval},
  title   = {A Survey of Query Auto Completion in Information Retrieval},
  year    = {2016},
  volume  = {10},
  issue   = {4},
  url     = {https://www.nowpublishers.com/article/Details/INR-055},
}

@Article{caldwell2011ethical,
  author    = {Caldwell, T.},
  Journal = {Network Security},
  title     = {Ethical hackers: putting on the white hat},
  year      = {2011},
  volume    = {2011},
  issue     = {7},
  url       = {https://www.sciencedirect.com/science/article/abs/pii/S1353485811700757},
}

@Online{Chandler-The-Sun-2018,
  author       = {S. Chandler},
  howpublished = {The Sun},
  title        = {Microsoft's Bing and Yahoo are showing users racist and 'ILLEGAL' paedo image results},
  year         = {2018},
  url          = {https://www.thesun.co.uk/tech/7467796/bing-racist-illegal-search-results},
}

@Online{chen2019gmail,
  author  = {Chen, M. X. and Lee, B. N. and Bansal, G. and Cao, Y. and Zhang, S. and Lu, J. and Tsay, J. and Wang, Y. and Dai, A. M. and Chen, Z. and Sohn, T. and Wu, Y.},
  howpublished = {arXiv.org},
  title   = {Gmail Smart Compose: Real-Time Assisted Writing},
  year    = {2019},
  url     = {https://arxiv.org/abs/1906.00080},
}

@InProceedings{cheng2017anyone,
  author    = {Cheng, J. and Bernstein, M. and Danescu-Niculescu-Mizil, C. and Leskovec, J.},
  booktitle = {Proc. of Computer Supported Cooperative Work},
  title     = {Anyone can become a troll: Causes of trolling behavior in online discussions},
  year      = {2017},
  url       = {https://dl.acm.org/doi/pdf/10.1145/2998181.2998213},
}

@TechReport{Cheung-SSRN-2015,
  author      = {Cheung, A. S. Y.},
  institution = {University of Hong Kong},
  title       = {Defaming by Suggestion: Searching for Search Engine Liability in the Autocomplete Era},
  year        = {2015},
  publisher   = {SSRN},
  url         = {https://papers.ssrn.com/sol3/papers.cfm?abstract_id=2611074},
}

@InProceedings{Chuklin-Sexi-2013,
  author    = {A. Chuklin and A. Lavrentyeva},
  booktitle = {Proc. of Workshop on Search and Exploration of X-rated Information},
  title     = {Adult query classification for web search and recommendation},
  year      = {2013},
  url       = {https://dl.acm.org/doi/pdf/10.1145/2433396.2433507},
}

@InProceedings{davidson2017automated,
  author    = {Davidson, T. and Warmsley, D. and Macy, M. and Weber, I.},
  booktitle = {Proc. of International Conference on Web and Social Media (ICWSM)},
  title     = {Automated hate speech detection and the problem of offensive language},
  year      = {2017},
  url       = {https://www.aaai.org/ocs/index.php/ICWSM/ICWSM17/paper/view/15665/14843},
}

@Online{diakopoulos2013sex,
  author       = {Diakopoulos, N.},
  howpublished = {Slate},
  title        = {Sex, violence, and autocomplete algorithms},
  year         = {2013},
  url          = {https://slate.com/technology/2013/08/words-banned-from-bing-and-googles-autocomplete-algorithms.html},
}

@online{diakopoulos2014algorithmic,
  author  = {Diakopoulos, N.},
  howpublished = {http://www.nickdiakopoulos.com/},
  title   = {Algorithmic defamation: the case of the shameless autocomplete},
  year    = {2013},
  url     = {http://www.nickdiakopoulos.com/2013/08/06/algorithmic-defamation-the-case-of-the-shameless-autocomplete/}
}

@TechReport{Diakopoulos-Columbia-2014,
  author    = {Diakopoulos, N.},
  title     = {Algorithmic Accountability Reporting: On the Investigation of Black Boxes},
  year      = {2014},
  publisher = {Columbia University Libraries},
  url       = {https://academiccommons.columbia.edu/doi/10.7916/D8ZK5TW2},
}

@Misc{wired2019,
  author       = {DiResta, R.},
  howpublished = {WIRED},
  title        = {The Complexity of Simply Searching for Medical Advice},
  year         = {2019},
  url          = {https://www.wired.com/story/the-complexity-of-simply-searching-for-medical-advice},
}

@Article{Elers-Te-Kahaoa-2014,
  author  = {S. Elers},
  journal = {Te Kaharoa},
  title   = {Maori Are Scum, Stupid, Lazy: Maori According to Google},
  year    = {2014},
  number  = {1},
  volume  = {7},
  url     = {https://ojs.aut.ac.nz/te-kaharoa/index.php/tekaharoa/article/view/45/44},
}

@Article{Ghatnekar-LLAELR-2013,
  author  = {Ghatnekar, S.},
  journal = {Loyola Law Review},
  title   = {Injury By Algorithm: A Look Into Google's Liability For Defamatory Autocompleted Search Suggestions},
  year    = {2013},
  volume  = {33},
  url     = {https://digitalcommons.lmu.edu/cgi/viewcontent.cgi?article=1581&context=elr},
}

@Misc{Guardian-2016,
  author       = {S. Gibbs},
  howpublished = {The Guardian},
  title        = {Google alters search autocomplete to remove 'are Jews evil' suggestion},
  year         = {2016},
  url          = {https://www.theguardian.com/technology/2016/dec/05/google-alters-search-autocomplete-remove-are-jews-evil-suggestion},
}

@Article{gitari2015lexicon,
  author  = {Gitari, N. D. and Zuping, Z. and Damien, H. and Long, J.},
  journal = {International Journal of Multimedia and Ubiquitous Engineering},
  title   = {A lexicon-based approach for hate speech detection},
  year    = {2015},
  number  = {4},
  volume  = {10},
  url     = {https://www.researchgate.net/profile/Damien_Hanyurwimfura2/publication/283125668_A_Lexicon-based_Approach_for_Hate_Speech_Detection/links/5b9ea001a6fdccd3cb5df826/A-Lexicon-based-Approach-for-Hate-Speech-Detection.pdf},
}

@Online{GLAAD-Guidelines,
  author       = {GLAAD},
  title        = {GLAAD Media Reference Guide - Lesbian / Gay / Bisexual Glossary Of Terms},
  url          = {https://www.glaad.org/reference/lgbtq},
  year         = {2011},
}

@Article{golebiewskid,
  author  = {Golebiewski, M., and {boyd}, {d.}},
  journal = {Data \& Society Research Institute},
  title   = {Data Voids: Where Missing Data Can Easily Be Exploited},
  year    = {2018},
  url     = {https://datasociety.net/wp-content/uploads/2018/05/Data_Society_Data_Voids_Final_3.pdf},
}

@Online{Bing-Autosuggest-Blog-2013b,
  author  = {A. Gulli},
  title   = {A Deeper Look at Autosuggest},
  url     = {https://blogs.bing.com/search/2013/03/25/a-deeper-look-at-autosuggest/},
  year    = {2013},
}

@InProceedings{Gupta-ECIR-2017,
  author = {P. Gupta and J. Santos},
  title = {Learning to Classify Inappropriate Query Completions},
  booktitle = {Proc. European Conference on Inforomation Retrieval},
  year = {2017},
  url = {https://pgupta.gitlab.io/pdf/gupta-ecir-2017.pdf},
}

@Online{gupta2018impersonation,
  author       = {Gupta, R. and Kondapally, R. and Kiran, C. R.},
  howpublished = {arXiv.org},
  title        = {Impersonation: Modeling Persona in Smart Responses to Email},
  year         = {2018},
  url          = {https://arxiv.org/abs/1806.04456},
}

@InProceedings{hagiwara2019teaspn,
  author    = {Hagiwara, M. and Ito, T. and Kuribayashi, T. and Suzuki, J. and Inui, K.},
  booktitle = {Proc. of Conf. on Empirical Methods in Natural Language Processing (EMNLP)},
  title     = {TEASPN: Framework and Protocol for Integrated Writing Assistance Environments},
  year      = {2019},
  url       = {https://www.aclweb.org/anthology/D19-3039.pdf}
}

@Online{Hoffman-How-To-Geek-2018,
  author       = {C. Hoffman},
  howpublished = {How to Geek},
  title        = {Bing Is Suggesting the Worst Things You Can Imagine},
  year         = {2018},
  url          = {https://www.howtogeek.com/367878/bing-is-suggesting-the-worst-things-you-can-imagine}
}

@Online{Holan-Politifact-2010,
  author       = {A. D. Holan},
  howpublished = {PolitiFact},
  title        = {Why do so many people think Obama is a Muslim?},
  year         = {2010},
  url          = {https://www.politifact.com/article/2010/aug/26/why-do-so-many-people-think-obama-muslim},
}

@Online{hosseini2017deceiving,
  author  = {Hosseini, H. and Kannan, S. and Zhang, B. and Poovendran, R.},
  howpublished = {arXiv.org},
  title   = {Deceiving Google's Perspective API Built for Detecting Toxic Comments},
  year    = {2017},
  url     = {https://arxiv.org/abs/1702.08138},
}

@InProceedings{Huang-CIKM-2013,
  author    = {Huang, P. and He, X. and Gao, J. and Deng, L. and Acero, A. and Heck, L.},
  booktitle = {Conference on Information and Knowledge Management (CIKM)},
  title     = {Learning Deep Structured Semantic Models for Web Search Using Clickthrough Data},
  year      = {2013},
  url       = {https://dl.acm.org/doi/pdf/10.1145/2505515.2505665},
}

@InProceedings{Hulden-EACL-2009,
  author    = {Hulden, M.},
  booktitle = {Proc. of Conf. of European Chapter of Association of Computational Linguistics (EACL)},
  title     = {Foma: A Finite-state Compiler and Library},
  year      = {2009},
  url       = {https://www.aclweb.org/anthology/E09-2008.pdf},
}

@Article{ibrahim2017facebook,
  author    = {Ibrahim, Y.},
  journal   = {Social Media + Society},
  title     = {Facebook and the Napalm Girl: Reframing the Iconic as Pornographic},
  year      = {2017},
  number    = {4},
  volume    = {3},
  publisher = {SAGE Publications},
  url       = {https://journals.sagepub.com/doi/pdf/10.1177/2056305117743140},
}

@Online{Jensen-NPR-2018,
  author       = {E. Jensen},
  howpublished = {NPR},
  title        = {NPR's Approach To A Reported Presidential Profanity Evolves},
  year         = {2018},
  url          = {https://www.npr.org/sections/publiceditor/2018/01/12/577631226/nprs-approach-to-a-reported-presidential-profanity-evolves},
}

@InProceedings{joslinmeasuring,
  author    = {Joslin, M. and Li, N. and Hao, S. and Xue, M. and Zhu, H.},
  booktitle = {Proc. of IEEE Symposium on Security and Privacy (SP)},
  title     = {Measuring and Analyzing Search Engine Poisoning of Linguistic Collisions},
  year      = {2019},
  url       = {https://personal.utdallas.edu/~shao/papers/joslin_sp2019.pdf},
}

@InProceedings{kannan2016smart,
  author = {Kannan, A. and Young, P. and Ramavajjala, V. and Kurach, K. and Ravi, S. and Kaufmann, T. and Tomkins, A. and Miklos, B. and Corrado, G. and Lukacs, L. and Ganea, M.},
  booktitle = {Proc. of ACM SIGKDD Conference on Knowledge Discovery and Data Mining (KDD)},
  title  = {Smart Reply: Automated Response Suggestion for Email},
  year   = {2016},
  url    = {https://arxiv.org/abs/1606.04870},
}

@Article{Karapapa_IJILT_2015,
  author  = {Karapapa, S. and Borghi, M.},
  journal = {International Journal of Law and Information Technology},
  title   = {Search Engine Liability for Autocomplete Suggestions: Personality, Privacy and the Power of the Algorithm},
  year    = {2015},
  number  = {3},
  volume  = {23},
  url     = {https://papers.ssrn.com/sol3/papers.cfm?abstract_id=2862667},
}

@InProceedings{kumar2017antisocial,
  author    = {Kumar, S. and Cheng, J. and Leskovec, J.},
  booktitle = {Proc. of International Conference on World Wide Web},
  title     = {Antisocial Behavior on the Web: Characterization and Detection},
  year      = {2017},
  url       = {https://dl.acm.org/doi/pdf/10.1145/3041021.3051106},
}

@Online{Lapowsky-Wired-2018,
  author       = {I. Lapowsky},
  howpublished = {WIRED},
  title        = {Google Autocomplete Still Makes Vile Suggestions},
  year         = {2018},
  url          = {https://www.wired.com/story/google-autocomplete-vile-suggestions},
}

@Misc{cnn2017,
  author       = {S. Larson},
  howpublished = {CNN},
  title        = {Offensive chat app responses highlight AI fails},
  year         = {2017},
  url          = {https://money.cnn.com/2017/10/25/technology/business/google-allo-facebook-m-offensive-responses/index.html},
}

@Article{Lee-IEEE-2002,
  author  = {Lee, P. Y. and Hui, S. C. and Fong, A. C. M.},
  journal = {IEEE Intelligent Systems},
  title   = {Neural Networks for Web Content Filtering},
  year    = {2002},
  volume  = {17},
  issue   = {5},
  url     = {https://ieeexplore.ieee.org/abstract/document/1039832},
}

@Online{magu2017detecting,
  author       = {Magu, R. and Joshi, K. and Luo, J.},
  howpublished = {arXiv},
  title        = {Detecting the Hate Code on Social Media},
  year         = {2017},
  url          = {https://arxiv.org/abs/1703.05443},
}

@Online{Bing-Autosuggest-Blog-2013a,
  author  = {D. Marantz},
  title   = {A Look at Autosuggest},
  url     = {https://blogs.bing.com/search/2013/02/20/a-look-at-autosuggest/},
  year    = {2013},
}

@Misc{medium2019,
  author       = {Scott, M. V},
  howpublished = {Medium},
  title        = {Even Grammarly Finds my Use of the Word Fat Offensive, But Why?},
  year         = {2019},
  url          = {https://medium.com/@michellevscott/even-grammarly-finds-my-use-of-the-word-fat-offensive-but-why-442b16e6dc7a},
}

@InProceedings{miller2017responsible,
  author    = {Miller, B. and Record, I.},
  booktitle = {New Media \& Society},
  title     = {Responsible epistemic technologies: A social-epistemological analysis of autocompleted web search},
  year      = {2017},
  url       = {https://journals.sagepub.com/doi/abs/10.1177/1461444816644805},
}

@InBook{Molek-Kozakowska-2010,
  author    = {K. Molek-Kozakowska},
  editor    = {Urszula Okulska and Piotr Cap},
  publisher = {J. Benjamins Publishing},
  title     = {Perspectives in Politics and Discourse},
  year      = {2010},
  series    = {Language Arts \& Disciplines},
}

@InProceedings{nobata2016abusive,
  author    = {Nobata, C. and Tetreault, J. and Thomas, A. and Mehdad, Y. and Chang, Y.},
  booktitle = {Proc. of International Conference of World Wide Web},
  title     = {Abusive language detection in online user content},
  year      = {2016},
  url       = {https://dl.acm.org/doi/pdf/10.1145/2872427.2883062}, 
}

@InProceedings{olteanu2018effect,
  author    = {Olteanu, A. and Castillo, C. and Boy, J. and Varshney, K. R.},
  booktitle = {Proc. of International AAAI Conference on Web and Social Media},
  title     = {The effect of extremist violence on hateful speech online},
  year      = {2018},
  url       = {https://www.aaai.org/ocs/index.php/ICWSM/ICWSM18/paper/view/17908/17013},
}

@Article{olteanu2019social,
  author    = {Olteanu, A. and Castillo, C. and Diaz, F. and Kiciman, E.},
  journal   = {Frontiers in Big Data},
  title     = {Social data: Biases, methodological pitfalls, and ethical boundaries},
  year      = {2019},
  pages     = {13},
  volume    = {2},
  publisher = {Frontiers},
  url       = {https://www.frontiersin.org/articles/10.3389/fdata.2019.00013/full},
}

@InProceedings{olteanu2017limits,
  author    = {Olteanu, A. and Talamadupula, K. and Varshney, K. R},
  booktitle = {Proc. of ACM Conference on Web Science (WebSci)},
  title     = {The limits of abstract evaluation metrics: The case of hate speech detection},
  year      = {2017},
  url       = {https://dl.acm.org/doi/pdf/10.1145/3091478.3098871},
}

@misc{Olteanu-2020,
  author = {A. Olteanu and F. Diaz and G. Kazai},
  title = {When Are Search Completion Suggestions Problematic?},
  year = 2020,
  howpublished = {Under submission}
}

@Article{palmer2001ethical,
  author    = {Palmer, C.},
  journal   = {IBM Systems J.},
  title     = {Ethical hacking},
  year      = {2001},
  number    = {3},
  volume    = {40},
  publisher = {IBM},
  url       = {http://www.security-science.com/pdf/ethical-hacking-by-palmer.pdf},
}

@Misc{parikh2012identifying,
  author    = {Parikh, J. and Suresh, B.},
  note      = {US Patent 8,280,871},
  title     = {Identifying offensive content using user click data},
  year      = {2012},
  publisher = {Google Patents},
  url       = {https://patentimages.storage.googleapis.com/d6/68/23/a9dd35cc281862/US8280871.pdf},
}

@InProceedings{robertson2019auditing,
  author    = {Robertson, R. E and Jiang, S. and Lazer, D. and Wilson, C.},
  booktitle = {Proc. of ACM Conference on Web Science (WebSci)},
  title     = {Auditing Autocomplete: Suggestion Networks and Recursive Algorithm Interrogation},
  year      = {2019},
  url       = {https://dl.acm.org/doi/pdf/10.1145/3292522.3326047},
}

@Misc{santos2017query,
  author    = {Santos, J. C. A. and Arnold, P. D. and Roper, W. F. and Gupta, P.},
  note      = {US Patent App. 15/174,188},
  title     = {Query classification for appropriateness},
  year      = {2017},
  publisher = {Google Patents},
  url       = {https://patentimages.storage.googleapis.com/14/39/f2/34a6f12452d2bb/US20170351951A1.pdf},
}

@InProceedings{schmidt2017survey,
  author  = {Schmidt, A. and Wiegand, M.},
  journal = {Proc. of Workshop on Natural Language Processing for Social Media},
  title   = {A Survey on Hate Speech Detection using Natural Language Processing},
  year    = {2017},
  url     = {https://www.aclweb.org/anthology/W17-1101.pdf},
}

@Article{sellars2016defining,
  author  = {Sellars, A.},
  journal = {Berkman Klein Center Research Publication},
  number  = {2016-20},
  title   = {Defining hate speech},
  year    = {2016},
  url     = {https://papers.ssrn.com/sol3/papers.cfm?abstract_id=2882244##},
}

@TechReport{Settles-2009,
  author    = {B. Settles},
  title     = {Active Learning Literature Survey},
  year      = {2009},
  publisher = {University of Wisconsin–Madison},
  url       = {http://digital.library.wisc.edu/1793/60660},
}

@InProceedings{Shen-WWW-2014,
  author    = {Shen, Y. and He, X. and Gao, J. and Deng, L. and Mesnil, G.},
  booktitle = {Proc. of International Conference on World Wide Web},
  title     = {Learning Semantic Representations Using Convolutional Neural Networks for Web Search},
  year      = {2014},
  url       = {https://dl.acm.org/doi/pdf/10.1145/2567948.2577348},
}

@InProceedings{shokouhi2013learning,
  author    = {Shokouhi, M.},
  booktitle = {Proc. of ACM SIGIR Conference on Research and Development in Information Retrieval},
  title     = {Learning to personalize query auto-completion},
  year      = {2013},
  url       = {https://dl.acm.org/doi/pdf/10.1145/2484028.2484076},
}

@InProceedings{silva2016analyzing,
  author    = {Silva, L. and Mondal, M. and Correa, D. and Benevenuto, F. and Weber, I.},
  booktitle = {Proc. of AAAI Conference on
Web and Social Media (ICWSM)},
  title     = {Analyzing the targets of hate in online social media},
  year      = {2016},
  url       ={https://www.aaai.org/ocs/index.php/ICWSM/ICWSM16/paper/viewFile/13147/12829},
}

@Online{Starling-13,
  author = {L. Starling},
  title  = {How to remove a word from Google autocomplete},
  url    = {http://www.laurenstarling.org/how-to-remove-a-word-from-google-autocomplete/},
  year   = {2013},
}

@Online{Google-Autocomplete-Blog-2018,
  author  = {D. Sullivan},
  title   = {How Google autocomplete works in Search},
  url     = {https://www.blog.google/products/search/how-google-autocomplete-works-search/},
  year    = {2018},
}

@Online{Tiku-Wired-2018,
  author       = {N. Tiku},
  howpublished = {WIRED},
  title        = {Most Republicans Think Tech Companies Support Liberal Views},
  year         = {2018},
  url          = {https://www.wired.com/story/the-partisan-divide-around-censorship-in-social-media/},
}

@Article{tripodi2018searching,
  author  = {Tripodi, F.},
  journal = {Data \& Society},
  title   = {Searching for alternative facts: Analyzing Scriptural Inference in conservative news practices},
  year    = {2018},
  url     = {https://datasociety.net/wp-content/uploads/2018/05/Data_Society_Searching-for-Alternative-Facts.pdf},
}

@Online{Project-Veritas-2019,
  author       = {{Project Veritas}},
  howpublished = {Project Veritas},
  title        = {Insider Blows Whistle \& Exec Reveals Google Plan to Prevent ''Trump Situation'' in 2020 on Hidden Cam},
  year         = {2019},
  url          = {https://www.projectveritas.com/news/insider-blows-whistle-exec-reveals-google-plan-to-prevent-trump-situation-in-2020-on-hidden-cam},
}

@Online{Vernon-2015,
  author       = {S. Vernon},
  howpublished = {Marketing Hy},
  title        = {How YouTube Autosuggest and Google Autocomplete Can Work In Your Favor},
  year         = {2015},
  url          = {https://marketinghy.com/2015/01/youtube-autosuggest-google-autocomplete-can-work-favor},
}

@Misc{theverge2018,
  author       = {J. Vincent},
  howpublished = {The Verge},
  title        = {Google removes gendered pronouns from Gmail's Smart Compose feature},
  year         = {2018},
  url          = {https://www.theverge.com/2018/11/27/18114127/google-gmail-smart-compose-ai-gender-bias-prounouns-removed},
}

@InProceedings{Wang-DSN-2013,
  author    = {Wang, G. and Stokes, J. W. and Herley, C. and Felstead, D.},
  booktitle = {Proc. of International Conference on Dependable Systems and Networks (DSN)},
  title     = {Detecting malicious landing pages in Malware Distribution Networks},
  year      = {2013},
  url       = {https://www.microsoft.com/en-us/research/wp-content/uploads/2016/02/dsn2013.pdf},
}

@InProceedings{Wang-NDSS-2018,
  author    = {P. Wang and X. Mi and X. Liao and X. Wang and K. Yuan and F. Qian and R. A. Beyah},
  booktitle = {Proc. of Network and Distributed Systems Security (NDSS)},
  title     = {Game of Missuggestions: {S}emantic Analysis of Search-Autocomplete Manipulations},
  year      = {2018},
  url       = {https://www.ndss-symposium.org/wp-content/uploads/2018/02/ndss2018_07A-1_Wang_paper.pdf},
}

@InProceedings{weber2010demographics,
  author    = {Weber, I. and Castillo, C.},
  booktitle = {Proc. of ACM SIGIR Conference on Research and Development in Information Retrieval},
  title     = {The demographics of web search},
  year      = {2010},
  url       = {https://dl.acm.org/doi/pdf/10.1145/1835449.1835537},
}

@Misc{Winterman-BBC-2008,
  author       = {D. Winterman},
  howpublished = {BBC},
  title        = {How 'gay' became children's insult of choice},
  year         = {2008},
  url          = {http://news.bbc.co.uk/2/hi/7289390.stm},
}

@Online{UN-Women-AD-Campaign-2013,
  author = {{UN Women}},
  title  = {UN Women ad series reveals widespread sexism},
  url    = {https://www.unwomen.org/en/news/stories/2013/10/women-should-ads},
  year   = {2013},
}

@Online{Wilson-Betanews-2016,
  author  = {M. Wyci\'{s}lik-Wilson},
  title   = {Google explains that search autocomplete censors suggestions},
  url     = {https://betanews.com/2016/06/11/google-censors-search-suggestions/},
  year    = {2016},
}

@Online{Yehoshua-Blog-2016,
  author  = {T. Yehoshua},
  title   = {Google Search Autocomplete},
  url     = {https://blog.google/products/search/google-search-autocomplete/},
  year    = {2016},
}

@InProceedings{Yenala-PAKDD-2017,
  author    = {H. Yenala and M. Chinnakotla and J. Goyal},
  booktitle = {Pacific-Asia Conference on Knowledge Discovery and Data Mining (PAKDD)},
  title     = {Convolutional Bi-directional LSTM for Detecting Inappropriate Query Suggestions in Web Search},
  year      = {2017},
  url       = {https://www.researchgate.net/profile/Manoj_Chinnakotla/publication/316369136_Convolutional_Bi-directional_LSTM_for_Detecting_Inappropriate_Query_Suggestions_in_Web_Search/links/5a0422c10f7e9beb1774ed18/Convolutional-Bi-directional-LSTM-for-Detecting-Inappropriate-Query-Suggestions-in-Web-Search.pdf},
}
